\documentclass[paper]{JHEP3}
\usepackage{amsmath}
\usepackage{bm}
\usepackage{graphicx}
\usepackage{cite}

\newcommand{\calM}{{\cal M}}
\newcommand{\calO}{{\cal O}}

\newcommand{\be}{\begin{equation}}
\newcommand{\ee}{\end{equation}}

\title{\begin{boldmath}A threshold-improved narrow-width approximation for BSM physics\end{boldmath}}

\author{N.~Kauer\thanks{e-mail: \texttt{kauer@physik.uni-wuerzburg.de}}\\
Institut f\"ur Theoretische Physik, Universit\"{a}t W\"{u}rzburg, D-97074 W\"{u}rzburg, Germany}

\abstract{A modified narrow-width approximation 
that allows for $\calO(\Gamma/M)$-accurate predictions 
for resonant particle decay with similar intermediate masses
is proposed and applied to MSSM processes to demonstrate its 
importance for searches for particle physics beyond the Standard Model.
}

\keywords{Beyond Standard Model, NLO Computations, Hadronic Colliders}

\preprint{}

\begin{document}


\section{Introduction}

Theoretical arguments and experimental observations indicate that new particles
or interactions play an important role at the TeV scale, which will become
directly accessible at the Large Hadron Collider (LHC) -- scheduled to start in 2008
-- and its planned complement, the International Linear Collider.
In the near future we can therefore anticipate ground-breaking discoveries that 
reveal physics beyond the Standard Model (BSM) and allow to gain insight into
the structure of the fundamental theory.  Theoretically appealing extensions
of the Standard Model (SM) often feature numerous additional interacting 
heavy particles.  Supersymmetric theories \cite{susy}, for example, are attractive,
because they solve the hierarchy problem and allow for the unification of
electroweak and strong interactions.  The Minimal Supersymmetric Standard Model (MSSM)
is one of the best studied candidates for BSM physics.
Its phenomenology is characterized by sparticle production and cascade decays,
which lead to many-particle final states and scattering amplitudes with complex 
resonance structure.  Cascade decays also occur in other extensions, e.g.~in 
universal extra dimensions models \cite{Macesanu:2005jx}.

In order to extract the additional Lagrangian parameters of an extended theory 
from collider data, theoretical predictions are required that match the 
experimental accuracies.  This can usually only be achieved by taking into 
account higher order corrections in perturbative calculations.
Next-to-leading order calculations for phenomenologically relevant
$2\to n$ processes with $n\gtrsim 4$ are technically very challenging or 
not yet feasible \cite[Sec.~30]{Buttar:2006zd}.  
Consequently, production and decay 
stages are regularly factorized by means of the narrow-width approximation (NWA),
which effectively results in on-shell intermediate states.\footnote{%
The NWA can thus not be applied if on-shell states are kinematically
forbidden.}
Its main advantage is that sub- and nonresonant as well as nonfactorizable 
amplitude contributions can be neglected in a theoretically consistent way.  
Huge calculational simplifications occur already at tree level.
For these reasons, the NWA is employed in nearly all studies of BSM physics.
Note that it is implicitly applied whenever branching ratios are extracted 
from scattering cross sections.  A reliable NWA uncertainty 
determination is therefore crucial.
Given the width $\Gamma$ and mass $M$ of an unstable 
particle, the uncertainty of the NWA is commonly estimated as 
$\calO(\Gamma/M)$ for each Breit-Wigner propagator that is integrated out,
with $\Gamma/M$ typically $\lesssim\ 5\%$.  For larger widths 
nonresonant contributions can no longer be neglected.

Recently, two circumstances have been observed in which the standard NWA
is not reliable \cite{Berdine:2007uv,Kauer:2007zc}: the first involves decays 
where a daughter mass $m$ approaches the parent mass $M$; the second involves 
the convolution of parton distribution functions with a resonant hard scattering 
process at center-of-mass energy $\sqrt{\hat{s}}$.
In this article we elucidate that both effects arise due to a significant 
deformation of the Breit-Wigner shape that is caused by threshold factors, 
and is not restricted to the region where the Breit-Wigner is cut off, 
i.e.~where $M-m$ or $\sqrt{\hat{s}}-M$ is approximately $\Gamma$.  
An essential factor is that the amplitude can contribute additional powers 
of the threshold factors, which strongly amplifies the effects.
For sample applications we then demonstrate that $\calO(\Gamma/M)$-accurate predictions can nevertheless be 
obtained by integrating out the Breit-Wigner in combination with the relevant 
threshold factors.


\section{\label{sec:modifications}NWA modifications}

To illustrate why the NWA error becomes unexpectedly large for mass configurations in 
an extended vicinity of kinematical bounds and how it can be modified in such cases, 
we consider the partial decay rate of a heavy 
particle $A$ that predominantly decays in two stages via an intermediate resonance 
$C$, i.e. $A\overset{1}{\to}B, C$ and $C\overset{2}{\to}D, E$.  
In terms of the $n$-body phase space element 
\be
d\phi(P;p_1,\dots,p_n) \equiv (2\pi)^4\delta^{(4)}(P-\sum_{i=1}^n p_i)\prod_{i=1}^n
\frac{d^3p_i}{(2\pi)^3 2 E_i}
\ee
and the matrix element $\calM$, the off-shell decay rate is given by
\begin{align}
\label{eq:ofs}
\Gamma_\text{off-shell} &=  \frac{1}{2M_A} \int d\phi |\calM|^2\\
&= \frac{1}{2M_A}\int\frac{dp_C^2}{2\pi}\; D(p_C^2) 
\int d\phi_1(p_C^2) \int d\phi_2(p_C^2)\; |\calM_r(p_C^2)|^2 \;.
\label{eq:gofsfac}
\end{align}
In Eq.~(\ref{eq:gofsfac}), the phase space factorization
\be
d\phi = d\phi_1 \frac{dp_C^2}{2\pi} d\phi_2
\ee
has been applied,
where $d\phi_1$($d\phi_2$) is the $2$-body phase space element of the 
first (second) decay 
stage.  In the rest frame of $A$,
\be
\label{eq:dphibeta}
d\phi_1 = \frac{1}{16\pi^2}\;\frac{|\bm{p}_C|}{M_A}\;d\Omega_C\ \ \text{with}\ \ |\bm{p}_C|=\frac{M_A}{2}\,\beta(M_B+\sqrt{p_C^2},M_A)\,\beta(M_B-\sqrt{p_C^2},M_A)\;,
\ee
where $\beta(m,M)\equiv \sqrt{1-m^2/M^2}$.  For $d\phi_2$ one finds an analogous 
expression.
In addition to $d\phi$, also $|\calM|^2$ has been factorized into the 
squared propagator denominator 
\be
D(p_C^2)\equiv \frac{1}{(p_C^2-M_C^2)^2+M_C^2\Gamma_C^2}
\ee
with 4-momentum $p_C$ and $|\calM_r|^2$, the residual squared amplitude for the 
$A\to B,D,E$ decay.  
Starting in Eq.~(\ref{eq:gofsfac}), we have indicated
the particularly relevant $p_C^2$-dependence of quantities explicitly, 
but suppressed the dependence on other kinematical variables.
In the limit $\Gamma_C\to 0$, $D(p_C^2)$ is asymptotically equal to 
$2\pi K\delta(p_C^2-M_C^2)$ with
\be
K=\frac{1}{2M_C\,\Gamma_C}=\int_{-\infty}^\infty \frac{dq^2}{2\pi}D(q^2)\;.
\ee
This replacement constitutes the standard NWA.
Employing it, one obtains
\begin{gather}
\label{eq:nwa}
\Gamma_\text{NWA} = \frac{1}{2M_A} K
\int d\phi_1(M_C^2) \int d\phi_2(M_C^2)\; |\calM_r(M_C^2)|^2 \;.
\end{gather}
It is suggestive to mitigate threshold-induced deviations by absorbing the amplifying 
factors of $\beta$-form that occur in $d\phi_1$ (see Eq.~(\ref{eq:dphibeta})) and 
$d\phi_2$ into $K$.  The absorbed factors have to be normalized to
$p_C^2=M_C^2$, so that they are not taken into account more than once.  
$K$ is thus replaced by
\begin{align}
\widetilde K = &\int_{(M_D+M_E)^2}^{(M_A - M_B)^2}\frac{dp_C^2}{2\pi}\ D(p_C^2) \notag\\
&\times \frac{\beta(M_B+\sqrt{p_C^2},M_A)\beta(M_B-\sqrt{p_C^2},M_A)}{\beta(M_B+M_C,M_A)\beta(M_B-M_C,M_A)} \notag\\ 
&\times \frac{\beta(M_D+M_E,\sqrt{p_C^2})\beta(M_D-M_E,\sqrt{p_C^2})}{\beta(M_D+M_E,M_C)\beta(M_D-M_E,M_C)} \notag\\ 
&\times \frac{f_{|\calM_r|^2}(\sqrt{p_C^2},M_A,M_B,M_D,M_E)}{f_{|\calM_r|^2}(M_C,M_A,M_B,M_D,M_E)}\,.\label{fullKINWA}
\end{align}
Below we find that additional amplifying factors 
like $M^2-m^2=\beta^2(m,M)M^2$ can arise due to momentum-dependent residual 
matrix elements.  Such factors are included generically in Eq.~(\ref{fullKINWA}) 
as $f_{|\calM_r|^2}$.  They are process specific and their powers depend on the 
spin/polarization of external states.

Note that replacing $K$ with $\widetilde K$ 
does not affect the intrinsic properties of the NWA 
and that it generalizes to multi-body decays.
A closed-form result for $R\equiv \widetilde K/K$ can only be given for special cases.
In Fig.~\ref{fig:Rdep} we show for the ratio $R_2$ that is obtained
by only taking into account the $\beta$-factors that arise from $d\phi_2$
and setting the upper integration boundary to infinity
the deviation from 1 normalized to $\Gamma_C/M_C$.  
One can see that the largest deviation occurs for $M_D+M_E\to M_C$
and $M_E$ (or $M_D$) $\to 0$.  
\FIGURE{
\begin{minipage}{0.7\textwidth}
\begin{center}
\includegraphics[height=7.5cm]{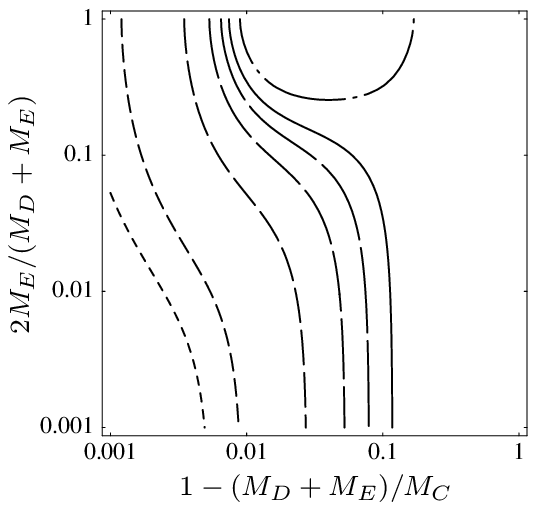}
\end{center}
\end{minipage}\\[0.2cm]
\caption{\label{fig:Rdep}The relative deviation of the modified from the standard NWA 
factor, normalized to $\Gamma_C/M_C = 1\%$, i.e.~$(R_2-1)/(\Gamma_C/M_C)$, is
shown as function of the masses of the second decay $C\to D,E$ 
(see main text for details).
Contour lines are shown for the values $0$ (solid), $-1$ (dot-dashed) and $1,3,10,50,100$ (dashed).  The dash length decreases with increasing magnitude.
\vspace*{0.4cm}}}
The sizable effect for $m\equiv M_D\approx M\equiv M_C$
and a small mass $M_E$, which we set to zero to obtain analytical results, 
is further amplified if the matrix element of the second decay contributes an
additional factor $M^2-m^2$.  This is for example the case if
$C$ and $E$ are fermions and $D$ is a scalar (and spin correlations between decay 1 and 2 are neglected) or if $C$ is a scalar and $D$ and
$E$ are fermions.  For these decay types, strong effects have been observed in 
Ref.~\cite{Berdine:2007uv} and Ref.~\cite{Kauer:2007zc}, respectively.
The corresponding $R_2'\equiv \widetilde K'_2/K$ is given by 
\begin{align}
&\left(\int_{m^2}^{q^2_\text{max}} \frac{dq^2}{2\pi}\frac{1}{(q^2-M^2)^2+(M\,\Gamma)^2}\,
\frac{(q^2-m^2)^2/q^2}{(M^2-m^2)^2/M^2}\right)\Bigg/\nonumber\\
&\left(\int_{-\infty}^\infty \frac{dq^2}{2\pi}\frac{1}{(q^2-M^2)^2+(M\,\Gamma)^2}
\right) \label{eq:integrands}
\end{align}
with $\Gamma\equiv\Gamma_C$.
In Fig.~\ref{fig:integrands} we show the deformation
of the Breit-Wigner shape due to the additional threshold factors
when $m$ approaches $M$.  
\FIGURE{
\begin{minipage}{0.7\textwidth}
\begin{center}
\includegraphics[height=6.5cm]{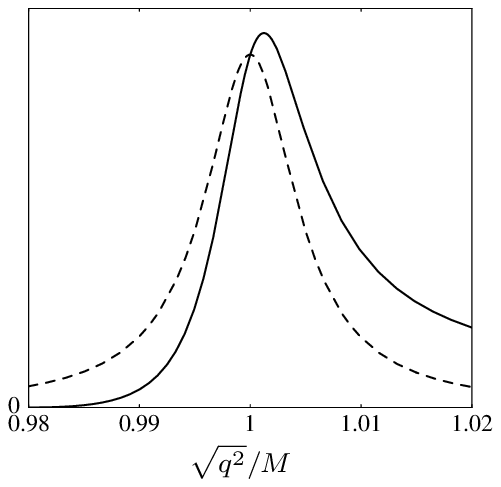}
\end{center}
\end{minipage}\\[0.2cm]
\caption{\label{fig:integrands}The Breit-Wigner shape deformation
is displayed that is caused by threshold factors when a decay daughter 
mass $m$ approaches the parent mass $M$.
More specifically, the integrand of the numerator (solid) and denominator (dashed) of 
Eq.~(\protect\ref{eq:integrands}) are shown in unspecified normalization 
as functions of the invariant mass $\sqrt{q^2}$.  $\Gamma/M=1\%$ and $m=M-2\Gamma$.
\vspace*{0.4cm}}}
After integration, we obtain
\begin{align}
R_2'=\,&\frac{1}{\pi}\left[\tan^{-1}\frac{\beta^2}{\gamma}+\tan^{-1}\frac{\lambda}{\gamma}\right]  \notag\\
&+\frac{\gamma}{\pi} \left[ 
\left(\frac{2}{\beta^2} - 1\right)\ln\frac{\lambda}{\beta^2} 
+\left(\frac{1}{\beta^2} - 1\right)^2\ln\frac{q^2_\text{max}}{m^2}\right]
\end{align}
with $\gamma\equiv\Gamma/M$, $\beta\equiv\beta(m,M)$ and $\lambda\equiv q^2_\text{max}/M^2-1$ when factors of $1+\gamma^2$ are approximated by $1$.
(This approximation does not produce a visible difference in Fig.~\ref{fig:betadep}.)
The result confirms that away from threshold, where $\beta\approx 1$, one obtains $R_2'\approx 1$ with $\gamma\ll \beta^2,\lambda$. When approaching the threshold, 
i.e.~$\beta\to 0$, the divergence of the second, formally $\gamma$-suppressed 
term overcompensates the decrease of the first term.
In Fig.~\ref{fig:betadep} we show the $\beta$ dependence of $R_2'$ for typical
values of $\sqrt{q^2_\text{max}}/M$.  The deviation from the standard NWA 
clearly exceeds $\calO(\Gamma/M)$ already for threshold masses $m$ that are 
still significantly below the resonant region roughly bounded by $M\pm\Gamma$.
\FIGURE{
\begin{minipage}{0.7\textwidth}
\vspace*{0.2cm}
\begin{center}
\includegraphics[height=7.7cm]{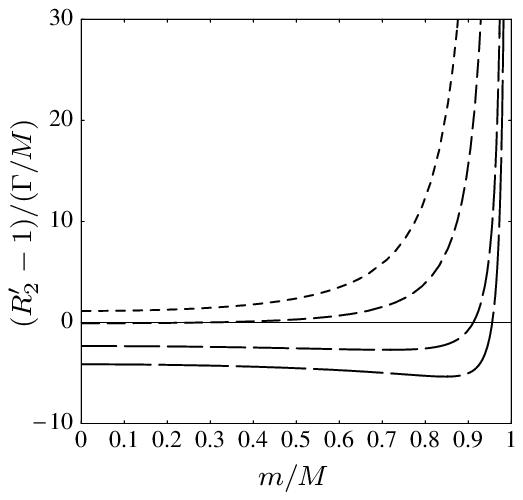}
\end{center}
\end{minipage}\\[0.2cm]
\caption{\label{fig:betadep}
The decay mass dependence of the relative deviation of the 
modified from the standard NWA factor is shown in units of the 
conventionally expected uncertainty 
$\Gamma/M=1\%$ for $\sqrt{q^2_\text{max}}/M\in\{1.05,1.1,2,10\}$.
The dash length decreases with increasing $\sqrt{q^2_\text{max}}$.} }
We note that if
$\sqrt{q^2_\text{max}}/M-1 \lesssim \gamma$ the threshold amplification is confined 
to the resonant region.  
However, in this case the arctan terms are no longer approximately $\pi/2$,
which results in a much larger than expected uncertainty of the standard NWA
for arbitrary values of $m$.  If $q^2_\text{max} \gg M^2$ the contribution
from the region $q^2\approx q^2_\text{max}$ to $R_2'$ is enhanced by the factor
$(q^2-m^2)^2/(q^2m^2)\approx q^2/m^2$ and the production cross 
section's suppression close to threshold can become important.


\section{\label{sec:applications}Applications}

In this section we demonstrate that the NWA 
modification proposed in Sec.~\ref{sec:modifications} allows to reduce the 
uncertainty to the conventional expectation for mass configurations in the 
vicinity of kinematical bounds.

As a first application we study the threshold-improved approximation for 
scalar scattering and decay.  
More specifically, we study the processes displayed in Fig.~\ref{fig:procs}.
For this type of process large standard NWA deviations have 
been observed in Refs.~\cite{Berdine:2007uv,Kauer:2007zc} when $m_d\approx M$
or the center of mass energy $\sqrt{s}\approx M$.
\FIGURE{
\begin{minipage}[c]{\linewidth}
\vspace*{0.4cm}
\begin{minipage}[c]{.49\linewidth}
\begin{center}
\includegraphics[height=2.7cm]{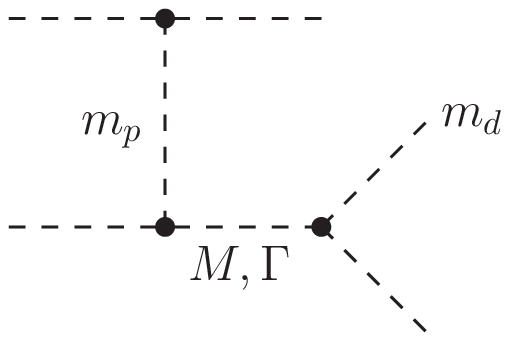}
\end{center}
\end{minipage} \hfill
\begin{minipage}[c]{.49\linewidth}
\vspace*{0cm}
\begin{center}
\includegraphics[height=2.7cm]{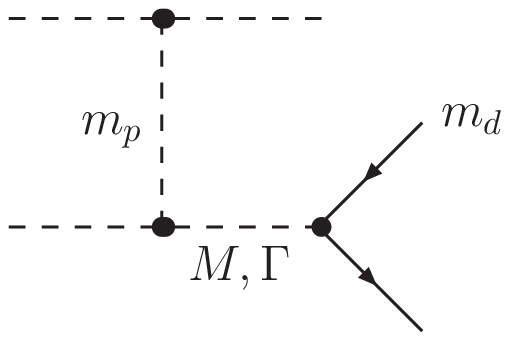}
\end{center}
\end{minipage}
\end{minipage}\\[0.cm]
\caption{{\label{fig:procs}
Process 1 (left) with scalar scattering and decay and process 2 (right) with
decay into non-scalar particles (fermions).
Lines without labels correspond to massless particles.}}}
We assess the quality of the modified NWA by comparing the off-shell cross section 
to the cross section 
$\sigma_\text{INWA}$ calculated in NWA with $\widetilde K$ of Eq.~(\ref{fullKINWA}),
where $M_A=\sqrt{s},M_C= M,\Gamma_C= \Gamma,M_D= m_d$ and 
$M_B=M_E= 0$.  The deviation is measured in units of $\Gamma/M$, which we set to
$0.01$, using
\be
R_\text{INWA} \equiv \left(\frac{\sigma_\text{off-shell}}{\sigma_\text{INWA}}-1\right)\bigg/\frac{\Gamma}{M}\;.
\ee
We start by neglecting matrix element effects and thus set $f_{|\calM_r|^2}=1$.
For process 1 with $m_p\sim M$ we find satisfactory NWA uncertainty reduction.
For instance for $m_p=1.1M$, $R_\text{INWA} \lesssim 3$ as long as 
$(\sqrt{s}-M)/M\gtrsim 10^{-5}$.  For $m_p\ll M$, however, large deviations
occur in a significant parameter space region, in particular for $m_d\approx M$.
For process 2 large deviations remain, independent of the value of $m_p$.
The Breit-Wigner deformation arises apparently not just from threshold-type 
phase space element factors.  To achieve a satisfactory NWA uncertainty reduction
it is in general essential to take into account factors originating from the
matrix element that distort the Breit-Wigner shape.
We separate them into production and decay-related factors:
\be
f_{|\calM_r|^2}(M,\sqrt{s},0,m_d,0,m_p) = f_p(M,\sqrt{s},m_p)\;f_d(M,m_d)\;.
\ee
For process 1, the decay matrix element is a coupling constant and we thus 
have $f_d=1$.  For process 2, however, we have
\be
f_d(M,m_d)=\frac{|\calM_d|^2}{m_d^2}=\frac{M^2-m_d^2}{m_d^2}=\beta^2(m_d,M)\frac{M^2}{m_d^2}\;,
\ee
where we have divided by $m_d^2$ to obtain a dimensionless quantity and
expressed the result in terms of threshold $\beta$-factors that also appear 
in the decay phase space element (see Eq.~(\ref{eq:dphibeta})) in order 
to show the deviation-amplifying effect of the decay matrix element by contributing 
an additional power to the Breit-Wigner deforming factor in $d\phi_d$.
When taking decay matrix element effects into account, i.e. employing $\widetilde K$
with $f_{|\calM_r|^2} = f_d(M,m_d)$, NWA deviations are mitigated to  
$\calO(\Gamma/M)$, except for the region $\sqrt{s}\lesssim 1.5M$,
where the $t$-channel production matrix element causes significant Breit-Wigner
deformations.  These production effects can be remedied with
\be
f_p(M,\sqrt{s},m_p) = \beta^{-2}\left(\sqrt{|M^2-m_p^2|},\sqrt{s}\right)\;.
\ee


We now extend our analysis to more complex processes and study the NWA uncertainty 
reduction at 
hadron colliders for sparticle production and decay in the MSSM.   
In Ref.~\cite{Berdine:2007uv} the standard NWA
accuracy was studied for the 
process $u\bar{d}\to (\tilde{g}\to \widetilde{s}_{L,R}\bar{s})\,\widetilde{\chi}_1^+$.
Here, the resonant particle, i.e.~the gluino, is produced in a $t$-channel process
with either $\tilde{d}_L$ or $\tilde{u}_L$ exchange.
For this process a variation of the $\widetilde{s}_{L}$ mass between $0$ and the 
gluino mass revealed unexpectedly large NWA deviations for squark masses that are 
larger than $0.8M_{\tilde{g}}$.  The slope of the increasing deviation when the
squark mass approaches the gluino mass is qualitatively very similar to the
slope displayed for $R_2'$ in Fig.~\ref{fig:betadep}.  Since $R_2'$ does not take 
into account the $t$-channel production effects, we conclude that they
do not significantly alter the dominant decay effects.
We have confirmed that in this region the uncertainty of the NWA
is reduced to $\calO(\Gamma/M)$ if 
$\widetilde K'_2$ of Eq.~(\ref{eq:integrands}) is used with 
$M=M_{\tilde{g}}$, $\Gamma=\Gamma_{\tilde{g}}$, $m=M_{\widetilde{s}}$ and 
$\sqrt{q^2_\text{max}}=1.4M_{\tilde{g}}$ (matched at the squark mass value where
$\sigma_\text{off-shell}=\sigma_\text{NWA}$).
We note that for $\widetilde{s}_L$ masses below $0.8M_{\tilde{g}}$ the NWA 
overestimates the off-shell cross section by up to about 20\%.  
This deviation is, however, consistent with an expected uncertainty of 
$\calO(\Gamma/M)$, since in this region the gluino width increases to about 
10\% of its mass.
Ref.~\cite{Berdine:2007uv} also illustrates deviations for the 
$\widetilde{s}_L$-$\widetilde{s}_R$ decay asymmetry, which are consistent with 
NWA corrections of $\calO(\Gamma/M)$.


As a last application we consider cascade decays, which are the natural testing 
ground for Eq.~(\ref{fullKINWA}).
More specifically, we study $\tilde{g}\,\widetilde{u}_L$
production at the LHC, i.e. in proton-proton collisions at 
14 TeV, with the subsequent cascade decay $\tilde{g} \to \widetilde{s}_L\bar{s}$ 
and $\widetilde{s}_L \to \widetilde{\chi}^-_1 c$ at the SPS1a' benchmark point 
\cite{sps1ap} in the MSSM parameter space.  Phenomenologically, to consider 
a squark decay into the LSP candidate $\widetilde{\chi}^0_1$ would be more natural,
but the resulting complete Feynman amplitude features a complicated resonance 
structure whose study we leave to future work.  
Even for the gluino decay chain considered here, interference arises from 
$\tilde{g} \to (\widetilde{c}^\ast_L \to \widetilde{\chi}^-_1 \bar{s}) c$.
Its effect is, however, small.  We confirmed that omitting it does not affect 
our $\calO(\Gamma/M)$ accuracy goal.
In this article we focus on the resonant $\widetilde{s}_L$ state 
(with $M=570$ GeV and $\Gamma=5.4$ GeV at SPS1a') and the NWA accuracy 
relative to $\Gamma/M$ that is obtained using
$\widetilde K$ with 
\be
f_{|\calM_r|^2}(M_{\widetilde{s}}, M_{\tilde{g}},0,M_{\widetilde{\chi}},0)=
\frac{M_{\tilde{g}}^2-M_{\widetilde{s}}^2}{M_{\tilde{g}}^2}\;\frac{M_{\widetilde{s}}^2-M_{\widetilde{\chi}}^2}{M_{\widetilde{\chi}}^2}=\beta^2(M_{\widetilde{s}},M_{\tilde{g}})\;\beta^2(M_{\widetilde{\chi}},M_{\widetilde{s}})\;\frac{M_{\widetilde{s}}^2}{M_{\widetilde{\chi}}^2}
\ee
versus $K$ when 
the strange squark mass approaches either the gluino or chargino mass of
$607$ and $184$ GeV, respectively.\footnote{The chargino is treated as stable
and the gluino in standard NWA with spin correlations.}
Results calculated with 
MadEvent \cite{madevent} and Sherpa \cite{sherpa} using CTEQ6L1 parton distribution 
functions \cite{Pumplin:2002vw} 
and spectra and decay widths obtained with SPheno \cite{Porod:2003um} and 
SDECAY \cite{Muhlleitner:2003vg} are displayed in Fig.~\ref{fig:mssm}.
\FIGURE{
\begin{minipage}[c]{\linewidth}
\vspace*{0.4cm}
\begin{minipage}[c]{.49\linewidth}
\flushleft \includegraphics[height=6.7cm]{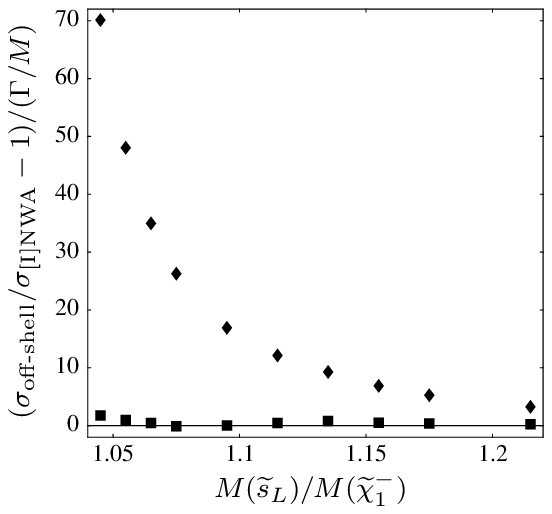}
\end{minipage} \hfill
\begin{minipage}[c]{.49\linewidth}
\vspace*{0.1cm}
\flushright \includegraphics[height=6.84cm]{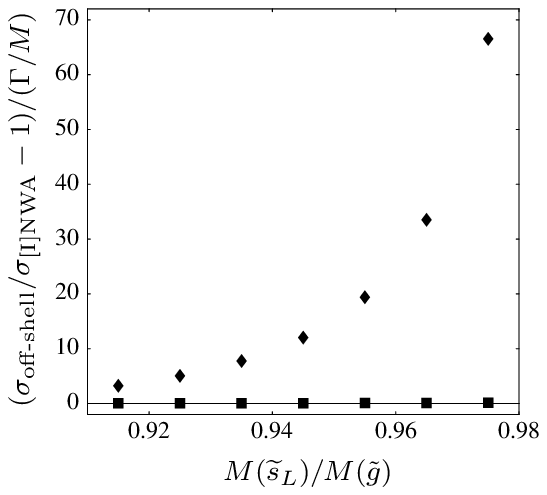}
\end{minipage}
\end{minipage}\\[0.2cm]
\caption{{\label{fig:mssm}
The accuracy of the NWA cross section normalized to the conventionally expected 
uncertainty is shown for $\tilde{g}\,\widetilde{u}_L$ production at the LHC followed 
by the cascade decay $\tilde{g} \to \widetilde{s}_L\bar{s}$ and 
$\widetilde{s}_L \to \widetilde{\chi}^-_1 c$ in the MSSM at SPS1a' for 
a variable strange squark mass that approaches the chargino mass (left) and 
the gluino mass (right).
Results are displayed for the standard NWA (diamonds) and the improved NWA (INWA) of 
Eq.~(\protect\ref{fullKINWA}) (boxes).  $\Gamma(\widetilde{s}_L)/M(\widetilde{s}_L)$ 
ranges from $0.03\%$ to $0.16\%$ (left) and is approximately $0.9\%$ (right).}}}
The Monte Carlo integration error is 0.1\%.
Both figures show that the modified NWA reduces 
the sizable deviations that occur in standard NWA as a daughter or parent mass is 
approached to the conventional uncertainty estimate.
A multiple, overlapping application of Eq.~(\ref{fullKINWA}) that would also include 
gluino production and chargino decay effects could be envisioned, but is beyond the 
scope of this work.


\section{Conclusions}

For configurations with kinematical bounds in the vicinity of resonances
phase space suppression via $\beta$-factors can significantly distort the
resonance Breit-Wigner, thus effecting an unexpectedly large NWA error.
For affected configurations we proposed a modification of the standard NWA that
allows to take this kinematical phase space suppression into account and thus 
to reduce the approximation uncertainty to the inverse of the generic resonant 
enhancement $M/\Gamma$.  For supersymmetric extensions of the SM 
we have demonstrated this uncertainty reduction for similar masses in processes with 
single particle or cascade decay.
If applied in phenomenological studies and data
analysis with tools like Fittino \cite{Bechtle:2004pc}, SFITTER \cite{Lafaye:2004cn} 
or MARMOSET \cite{ArkaniHamed:2007fw}, 
the method would contribute to an accurate determination 
of BSM model parameters and thus to establishing supersymmetry or other key 
properties of the fundamental theory.


\acknowledgments
I would like to thank the Galileo Galilei Institute for Theoretical
Physics for the hospitality and the INFN for partial support during the
completion of this work.
This work was supported by the BMBF, Germany (contract 05HT1WWA2).


\end{document}